\documentclass{article}

\usepackage{microtype}
\usepackage{graphicx}
\usepackage{subfigure}
\usepackage{booktabs} 

\usepackage{hyperref}

\usepackage[accepted]{ml4astro2025}

\usepackage{amsmath}
\usepackage{amssymb}
\usepackage{mathtools}
\usepackage{amsthm}
\usepackage{longtable} 
\usepackage{pdflscape}
\usepackage{gensymb}
\usepackage[capitalize,noabbrev]{cleveref}
\usepackage{xcolor}   
\usepackage{tikz}
\usepackage{fontawesome}
\usepackage{dashrule}
\definecolor{mpltab:blue}{HTML}{1f77b4}
\definecolor{mpltab:orange}{HTML}{ff7f0e}
\definecolor{mpltab:green}{HTML}{2ca02c}
\definecolor{mpltab:red}{HTML}{d62728}
\definecolor{mpltab:purple}{HTML}{9467bd}

\theoremstyle{plain}

\theoremstyle{definition}

\theoremstyle{remark}

\usepackage[textsize=tiny]{todonotes}

\mlforastrotitlerunning{Causal Evidence for the Primordiality of Colors in Trans-Neptunian Objects}

\begin{document}

\twocolumn[
\mlforastrotitle{Causal Evidence for the Primordiality of Colors in Trans-Neptunian Objects}



\mlforastrosetsymbol{equal}{*}

\begin{mlforastroauthorlist}
\mlforastroauthor{Benjamin L.\ Davis*}{nyuad}
\mlforastroauthor{Mohamad Ali-Dib*}{nyuad}
\mlforastroauthor{Yujia Zheng*}{cmu}
\mlforastroauthor{Zehao Jin*}{nyuad,fudan}
\mlforastroauthor{Kun Zhang}{cmu,mbz}
\mlforastroauthor{Andrea Valerio Macci\`{o}}{nyuad}
\end{mlforastroauthorlist}

\mlforastroaffiliation{nyuad}{Center for Astrophysics and Space Science (CASS), New York University Abu Dhabi, PO Box 129188, Abu Dhabi, UAE}
\mlforastroaffiliation{cmu}{Carnegie Mellon University, Pittsburgh, PA, USA}
\mlforastroaffiliation{fudan}{Center for Astronomy and Astrophysics and Department of Physics, Fudan University, Shanghai 200438, People’s Republic of China}
\mlforastroaffiliation{mbz}{Mohamed bin Zayed University of Artificial Intelligence, Abu Dhabi, UAE}

\mlforastrocorrespondingauthor{Benjamin L.\ Davis}{ben.davis@nyu.edu}

\mlforastrokeywords{Scientific Discovery, Physics}

\vskip 0.3in
]



\printAffiliationsAndNotice{\mlforastroEqualContribution} 

\begin{abstract}
The origins of the colors of Trans-Neptunian Objects (TNOs) represent a crucial unresolved question, central to understanding the history of our Solar System.
Recent observational surveys have revealed correlations between the eccentricity and inclination of TNOs and their colors.
This has rekindled the long-standing debate on whether these colors reflect the conditions of TNO formation or their subsequent collisional evolution.
In this study, we address this question with 98.7\% certainty, using a model-agnostic, data-driven approach based on causal graphs.
First, as a sanity check, we demonstrate how our model can replicate the currently accepted paradigms of TNOs' dynamical history, blindly and without any orbital modeling or physics-based assumptions.
In fact, our causal model (with no knowledge of the existence of Neptune) predicts the existence of an unknown perturbing body, i.e., Neptune.
We then show how this model predicts, with high certainty, that the color of TNOs is the root cause of their inclination distribution, rather than the other way around. 
This strongly suggests that the colors of TNOs reflect an underlying dynamical property, most likely their formation location.
Moreover, our causal model excludes formation scenarios that invoke substantial color modification by subsequent irradiation.
We therefore conclude that the colors of TNOs are predominantly primordial.
\end{abstract}

\section{Introduction}\label{sec:intro}

Trans-Neptunian Objects (TNOs) are invaluable probes into the history and evolution of our Solar System \citep{morby1}. 
However, the wealth of information they encode is often difficult to decipher.
This includes intrinsic characteristics such as their sizes and correlated properties such as their orbits and surface photometric colors.
The last two have long been closely examined in an effort to unravel the relation between them \citep{Jewitt:2001}.

Although the history of these studies is long, here we focus on \citet{Marsset:2019} who found a strong correlation between the inclination and colors of TNOs. These results were expanded by \citet{Ali-Dib:2021}, who found an analogous trend where the eccentricity of VROs is cutoff at 0.42. They concluded that, in causality theory jargon, eccentricity ($e$) and inclination ($i$) are caused by the colors, which are indicative of the formation location. The primordial origin hypothesis of the TNO color diversity argues that TNO colors reflect compositional gradients in the protoplanetary disk, preserved since formation \citep{nesvorny2020,Ali-Dib:2021}. However, in alternatively, many works \citep{Luu:1996b,Stern:2002} argued that collisional evolution is the origin of TNO colors, where collisions expose fresh subsurface ices or organic materials, altering albedo and spectral slopes. A third possibility proposed that initially diverse bulk compositions undergo selective volatile evaporation post-formation, establishing steep compositional gradients across the primordial disk that, coupled with subsequent UV photolysis and particle irradiation, yield distinct surface chemistries \citep{brown1,wongbrown}.

In this paper, we use a purely data-driven, model-agnostic, statistical causal discovery method to study the relationships between the dynamical parameters of TNOs, and between those and the TNO colors.
We show that not only this technique allows us to derive some of the main lines of the current consensus on the origins of TNOs, but also that it elucidates the direction of causality between the dynamical parameters and colors of TNOs, and predicts the existence of an unknown perturbing body, i.e., Neptune.

\section{Methodology}\label{sec:methods}

\subsection{Data}\label{sec:data}

Our dataset is based on (but not exclusively) the Col-OSSOS survey \citep{Schwamb:2019}.
It was taken from \citet{Marsset:2019} and \citet{Ali-Dib:2021}.
It consists of a total of 229 TNOs including hot classicals, centaurs, and resonant/scattered objects, in a dataset for which discovery biases were modeled.
For each TNO, we have three orbital elements: semimajor axis ($a$), eccentricity ($e$), and inclination ($i$); and we have spectral slope (i.e., color).

\begin{figure*}
\centering
\includegraphics[clip=true, trim= 4mm 3mm 3mm 4mm, scale=0.4]{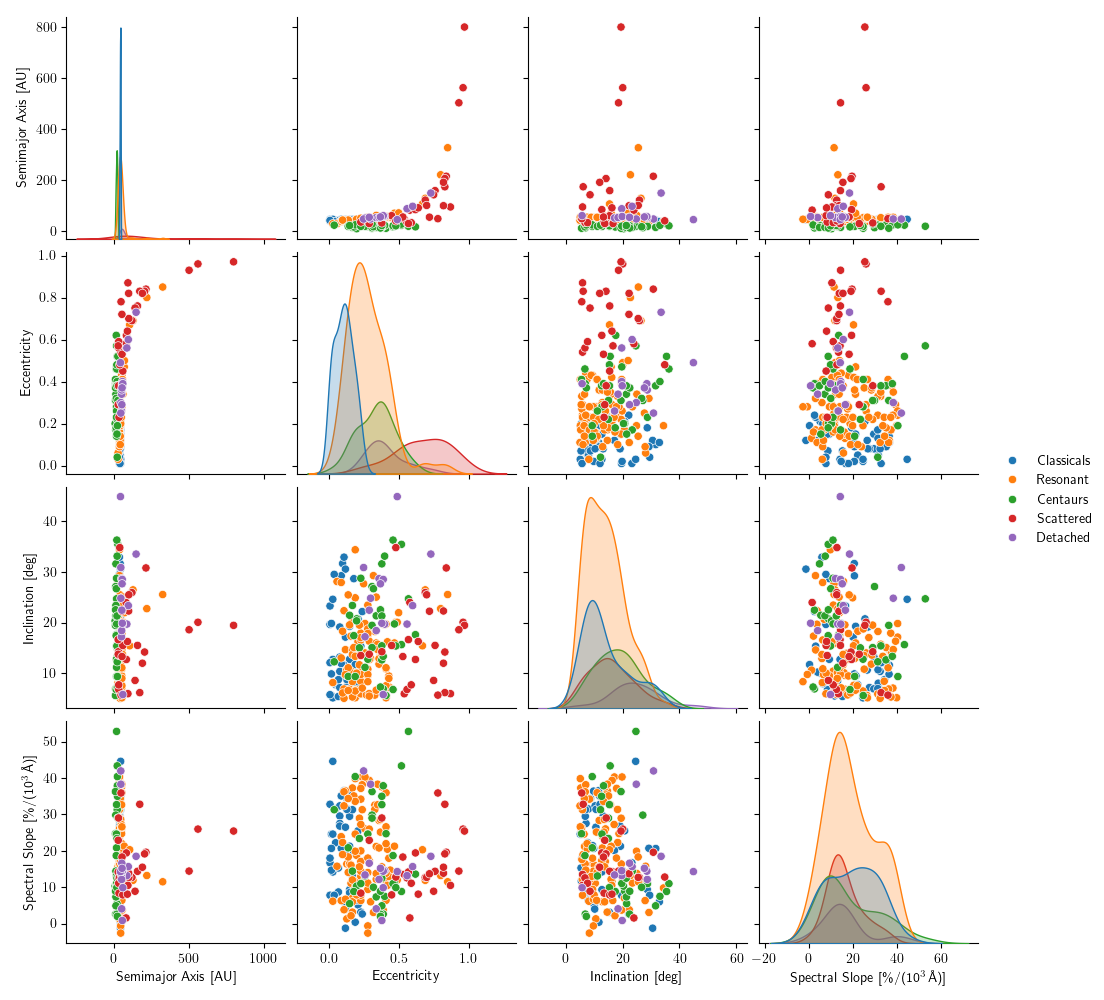}
\caption{Pairplot of all 229 TNOs in our study.
The TNOs are further divided into their individual populations: 48 Classicals (\textcolor{mpltab:blue}{$\bullet$}), 102 Resonant (\textcolor{mpltab:orange}{$\bullet$}), 36 Centaurs (\textcolor{mpltab:green}{$\bullet$}), 28 Scattered (\textcolor{mpltab:red}{$\bullet$}), and 15 Detached (\textcolor{mpltab:purple}{$\bullet$}).}
\label{fig:pairplot}
\end{figure*}

A fundamental assumption of this work is that colors are primordial, and thus strongly correlated to the initial location of a TNO.
Hereafter, we treat colors as a proxy for the initial semimajor axis of the objects. 
See Fig.~\ref{fig:pairplot} for a pairplot showing all the pairwise relations between our data.
Additionally, Fig.~\ref{fig:pairplot} shows the subpopulation in our data by separating each TNO by its classifications as either a Classical (48), Resonant (102), Centaur (36), Scattered (28), or Detached (15) object. 

\subsection{Causal Discovery: Identifying Cause-effect Relationships}

Identifying cause-effect relationships is crucial for moving beyond mere correlation to uncover the underlying causal mechanisms governing a system.
Traditionally, causal relationships are established through interventions or randomized experiments, where one variable is explicitly manipulated while all others are held constant, and the resulting effects are observed.
However, such interventions are infeasible in fields like astronomy, where the ``test subjects'' exist at unreachable astronomical distances.
Consequently, advanced methods are required to infer causal relationships from purely observational data---an endeavor that lies at the core of causal discovery \citep{spirtes2000causation}.

The foundation of causal discovery lies in uncovering the footprints of causality embedded in data.
One of the most important sources of such information is dependency relations.
By analyzing conditional independence among different components of an observed system, we can infer causal relationships between pairs of variables.
This allows us to construct a graph that encodes the results of essential conditional independence tests, revealing which variables \emph{cause} others under appropriate conditions.
Ideally, the output is a Directed Acyclic Graph (DAG) for a unique solution or a Completed Partially Directed Acyclic Graph (CPDAG) for a Markov equivalence class.
However, when some variables remain unmeasured, certain causal relationships may be undetermined, leading to a Partial Ancestral Graph (PAG).
For further reading on causal discovery and causality, see \textit{Causation, Prediction, and Search} \citep{spirtes2000causation}, \textit{Causality} \citep{pearl2009causality}, or the review in \citet[][\S2]{Jin:2025}.\footnote{
In addition to \citet{Jin:2025}, readers may be interested in further applications of causal discovery to astrophysical data \citep{Pasquato:2023,Pasquato:2024,Jin:2024,Jin:2025AAS}.
}

\subsection{Causal Structures with Latent Variables}

Since it is impossible to measure all variables in the Universe, latent variables are always present.
These unmeasured variables can significantly impact the correctness of the causal structure discovered.
For example, suppose that $X$ and $Y$ are independent in the general population, but a sample is selected based on a variable $Z$ that influences both $X$ and $Y$.
In that case, $X$ and $Y$ may exhibit statistical dependence in the sample, even though no such relationship exists in the population.
This can lead to spurious causal conclusions, falsely suggesting a direct causal relationship between $X$ and $Y$.

To address this challenge, we employ a principled approach capable of uncovering causal relationships even in the presence of latent variables.
A widely used method for this purpose is Fast Causal Inference \citep[FCI;][]{spirtes1995causal, zhang2008completeness}, a constraint-based algorithm that has been proven to provide sound causal conclusions despite unmeasured variables.
FCI has been applied across various scientific domains, including biology, economics, and climate science.
For our analysis of TNO orbits, we use the FCI implementation in the \texttt{Python} package \texttt{causal-learn} \citep{zheng2024causal} to infer the underlying causal structure.

FCI discovers causal relationships by performing a series of conditional independence (CI) tests.
These tests examine whether the statistical dependence between two variables disappears when controlling for other variables.
If two variables become independent when conditioning on a third, this suggests that the third variable may be an intermediary or a common cause.

Unlike many causal discovery methods that assume that all relevant variables are measured (such as those producing DAGs or CPDAGs), FCI accounts for the possibility of unobserved variables.
As a result, its output is a PAG, which provides more nuanced causal information.
The edges in a PAG have different interpretations:
\begin{itemize}
    \item $X \xrightarrow{\hspace{4.5mm}} Y $: $X$ is a \textit{cause} of $Y$.
    \item $X\ \circ\hspace{-1.78mm}\longrightarrow Y$: $Y$ is not an \textit{ancestor} of $X$.
    Intuitively, this implies $Y$ cannot be a cause of $X$, whether directly or indirectly.
    \item $X\ \circ$---$\circ \  Y$: No set d-separates $X$ and $Y$. In other words, they may be causally adjacent or share a latent common cause.
    \item $X\longleftrightarrow Y$: There is a latent common cause of $X$ and $Y$.
\end{itemize}
Therefore, by accounting for latent variables in the discovery process, we can uncover causal relations among measured variables while acknowledging uncertainties introduced by unmeasured factors.
More importantly, when the algorithm cannot determine a definitive causal direction due to latent variables, it explicitly represents this uncertainty rather than arbitrarily assigning a direction.
This principled approach distinguishes causal analysis from correlation-based techniques, ensuring that conclusions are drawn with a clear acknowledgment of underlying assumptions and limitations.

\section{Results}\label{sec:results}
\subsection{Data-driven Results}

\begin{figure}
    \centering
    \includegraphics[width=0.48\linewidth]{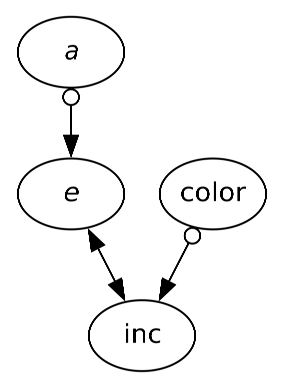}
    \caption{
    Partial Ancestral Graph (PAG) for 229 TNOs, calculated with the Fast Causal Inference (FCI) algorithm \citep{SpirtesManuscript-SPIAAA,Spirtes:2013,zheng2024causal}, for linear Fisher-Z conditional independence tests \citep{fisher_probable_1921} on transformed data, with $\alpha=0.013$ (significance level of individual partial correlation tests).
    On untransformed data, we recover the identical PAG with linear Fisher-Z tests and $\alpha=0.02$, while the same PAG is produced with $\alpha=0.09$ when we run Kernel-based conditional independence (KCI) tests \citep{Zhang:2012}, with a polynomial kernel.
    This PAG has three causal edges, which can be described as follows: (i) eccentricity is not an ancestor of the semimajor axis, (ii) there is a latent common cause of eccentricity and inclination, and (iii) inclination is not an ancestor of color.
    }
    \label{fig:fci}
\end{figure}

Our primary findings are summarized in Fig.~\ref{fig:fci} showing the statistically most likely PAG fitting our data, at 98.7\% confidence.
This main result utilizes the FCI algorithm \citep{SpirtesManuscript-SPIAAA,Spirtes:2013,zheng2024causal}, with linear Fisher-Z conditional independence tests \citep{fisher_probable_1921}, and the threshold for each conditional independence test is $\alpha=0.013$ (i.e., all tests must pass at the 98.7\% level) on transformed data via Gaussianization.
Specifically, we employ the Yeo-Johnson transformations \citep{Yeo:2000} for our primary preprocessing.
We still get the same PAG using a linear Fisher-Z test without any transformation for $\alpha=0.02$.
It is also possible to directly use a non-linear conditional independence test.
Here, we adopt a Kernel-based conditional independence (KCI) test \citep{Zhang:2012}, with a polynomial kernel and reproduce the same PAG as in Fig.~\ref{fig:fci} at $\alpha=0.09$.\footnote{We did not apply non-linear tests in the first place because a non-linear method is prone to overfitting for the relatively small size of our data (229 TNOs).}

We emphasize that this PAG was obtained with a purely data-driven approach, without astrophysical insights.
Moreover, we consistently reproduce the same PAG as in Fig.~\ref{fig:fci} by jackknifing our data by sequentially leaving out each subpopulation of TNOs.
Thus, removing any subsample of 48 Classicals, 102 Resonant, 36 Centaurs, 28 Scattered, or 15 Detached TNOs results in no change to our discovered PAG.
Therefore, we demonstrate that no single subpopulation is dominating the PAG and that our results are robust to outliers.

Alternatively, if we are to generate PAGs for the individual populations separately (i.e., analyzing only one subpopulation at a time), we find a large diversity in the results.
Many of these PAGs however are based on very few datapoints.
Taking this result at face value hints that our overall PAG represents that main-line dynamics dominate over the entire sample.

\subsection{Astrophysical Interpretation}

\textbf{The first link} we investigate is the one-way causal direction of the \textit{current} semimajor axis causing the \textit{current} eccentricity.
While the correlation between $a$ and $e$ in TNOs is well established, the direction of the causality we find here is not surprising either, as its root physical causes are:
\begin{itemize}
    \item Scattering by Neptune, where objects have to close-encounter Neptune first in order to get scattered into high eccentricity orbits. 
    Moreover, objects usually cannot be both close to Neptune (today) and have a high eccentricity.
    It is the current semimajor axis of the objects that dictates what eccentricity they can have, and not the other way around. 
    \item Mean motion resonances (MMRs), where the period (and thus current semimajor axis) of the objects dictates whether they are inside an eccentricity-raising resonance. 
\end{itemize}
The connection $a\ \circ\hspace{-1.2mm}\rightarrow e$ \emph{rules out} the possibility of $a\leftarrow e$.
Clearly, $a\rightarrow e$ is possible, but also $a\leftrightarrow e$.
The latter might imply that an unobserved confounder causes both $a$ and $e$.

\textbf{The second link} in the PAG is the two-way dependency between the eccentricity and the inclination, which is consistent with the von Zeipel-Lidov-Kozai \citep{Zeipel:1910,Lidov:1962,Kozai:1962} anti-correlated oscillations between these two quantities (both inside and outside of MMRs), that plays a central role in the dynamics of TNOs.
Here, $e\leftrightarrow i$ implies that there is an unobserved confounder.
Indeed, the von Zeipel-Lidov-Kozai mechanism involves perturbations from a third body, here being Neptune.
Moreover, if Neptune had not already been discovered in 1846, our result here would strongly suggest the presence of an unknown perturbing body.
Together, the first two links successfully reestablish the main dynamical processes shaping the Kuiper belt (scattering, MMRs, and von Zeipel-Lidov-Kozai oscillations) without any physical inputs.  

Finally, \textbf{the third piece} of the puzzle is the connection $\textrm{color}\ \circ\hspace{-1.2mm}\rightarrow i$ \emph{ruling out} the possibility of $\textrm{color}\leftarrow i$.
The ``color'' (i.e., a proxy for the formation location in our null hypothesis) is hence causing the inclination.
This is again dynamically expected, as the formation location relative to inclination-raising secular resonances such as f$_7$ and f$_8$  will strongly affect the inclination distribution of TNOs \citep{orangebook}.
Note that this link, however, leaves open the possibility of an unobserved confounder causing both color and the inclination.
This confounder can be the formation location itself, if we were to assume the color and initial location to be two distinct variables instead of the color being a proxy for location. 

Our result, that $\textrm{color}\leftarrow i$ is not allowed, rules out the model of \citet{Luu:1996b} and \citet{Stern:2002}, where collisional evolution shapes the colors of TNOs.
Moreover, our result that $\textrm{color}\leftarrow a$ is not allowed either, rules out the model of \citet{brown1} and \citet{wongbrown}, where $a$ would control the amount of irradiation a TNO is subjected to.
We are hence left only with the ``primordial origins'' model where \textbf{the color is set entirely by the chemical composition of the formation location}.

\section{Discussion \& Conclusions}\label{sec:discussion}

This work endeavors to resolve the tension between theories of primordial origins vs.\ subsequent evolution to account for the observed dispersion and correlations in TNO colors, a subject of a long debate. 
Our causal graph analysis, derived from a model-agnostic causal discovery framework, strongly favors the primordial origin hypothesis, with 98.7\% certainty that \emph{TNO color is causally antecedent to inclination, not a consequence of it.}
Moreover, our model seems to exclude any effects from the current semimajor axis on the color of TNOs, disfavoring models where continuous irradiation plays a large role in shaping the colors.
This will be explored further in the future. 
Finally, this work is a proof of principle for the use of causality models in planetary sciences.
In our case, it even ``rediscovers'' the existence of Neptune.
With large datasets ranging from asteroids to exoplanets, many discoveries await.

\section*{Acknowledgements}
This material is based on work supported by Tamkeen under the NYU Abu Dhabi Research Institute grant CASS.
YZ and KZ are supported by NSF Award No.~2229881, AI Institute for Societal Decision Making, NIH R01HL159805, and grants from Quris AI, Florin Court Capital, and MBZUAI-WIS Joint Program.
This research was carried out on the high-performance computing resources at New York University Abu Dhabi.
This research has made use of NASA's Astrophysics Data System Bibliographic Services.
The data and code used for this work are available for download from the following GitHub repository: \href{https://github.com/ZehaoJin/causalTNOs}{\faGithub~\url{https://github.com/ZehaoJin/causalTNOs}}.

\section*{Impact Statement}
This paper presents work whose goal is to advance the field of Machine Learning.
There are many potential societal consequences of our work, none which we feel must be specifically highlighted here.

\bibliography{bibliography}{}
\bibliographystyle{icml2025}

\newpage
\appendix
\onecolumn

\end{document}